\documentclass[preprintnumbers,amsmath,11pt,amssymb,floatfix,
superscriptaddress,nofootinbib,showpacs]{revtex4}
\usepackage{graphicx}
\usepackage{epsfig}
\usepackage{bm}
\usepackage{amsfonts}

\input epsf

\begin{document}

\title{\Large Interacting Ricci Dark Energy and its Statefinder Description}

\author{\bf Surajit Chattopadhyay}
\email{surajit_2008@yahoo.co.in,
surajit.chattopadhyay@rediffmail.com} \affiliation{Pailan College
of Management and Technology, Bengal Pailan Park, Kolkata-700 104,
India.}

\date{\today}

\begin{abstract}
In this paper we have considered an interacting Ricci dark energy
in flat FRW universe. We have reconstructed the Hubble's parameter
under this interaction. Also, we have investigated the statefinder
diagnostics. It has been revealed that the equation of state
parameter behaves like quintessence in this interaction and from
the statefinder diagnostics it has been concluded that the
interacting Ricci dark energy interpolates between dust and
$\Lambda$CDM stages of the universe.
\end{abstract}

\pacs{98.80.-k, 95.36.+x}

\maketitle

\subsection{\bf{Introduction}}
The ``dark energy" that is responsible for the present accelerated
expansion of the universe occupies about $70\%$ of today's
universe. Reviews on DE include \cite{saridakis}, \cite{sami},
\cite{padmanabhan} and \cite{Li}. The basic characteristic of DE
is that its equation of state (EOS) parameter $w =p/\rho$, where
$\rho$ is the energy density and $p$ is the pressure that has a
negative value. The simplest candidate of dark energy is a tiny
positive cosmological constant \cite{sami} corresponds to a fluid
with a constant equation of state $w=-1$. However, as is well
known, it is plagued by the so-called ``cosmological constant
problem" and ``coincidence problem" \cite{sami}. Other dark energy
models include quintessence \cite{Ratra}, phantom \cite{Najiri},
quintom \cite{saridakis}, Chaplygin gas \cite{Gorini}, tachyon
\cite{chimento}, hessence \cite{Zhao} etc. All DE models can be
classified by the behaviors of equations of state as following
\cite{saridakis}:
\begin{itemize}
    \item Cosmological constant: its EoS is exactly equal to $-1$, that is $w_{DE}=-1$.
    \item Quintessence: its EoS remains above the cosmological constant boundary, that is $w_{DE}\geq-1$.
    \item Phantom: its EoS lies below the cosmological constant boundary, that is $w_{DE}\leq-1$.
    \item Quintom: its EoS is able to evolve across the cosmological constant boundary.
\end{itemize}
In recent times, an interesting attempt for probing the nature of
dark energy within the framework of quantum gravity is the
so-called ''holographic dark energy'' (HDE) proposal \cite{Hsu}.
The holographic principle is an important result of the recent
research for exploring the quantum gravity \cite{Xhang}. Inspired
by the holographic principle, Gao et al.\cite{Gao} took the Ricci
scalar as the IR cut-off and named it the Ricci dark energy (RDE),
in which they take the Ricci scalar $R$ as the IR cutoff. With
proper choice of parameters the equation of state crosses $-1$, so
it is a `quintom' \cite{Feng}. The Ricci scalar of FRW universe is
given by $R=-6\left(\dot{H} +2H^{2}+\frac{k}{a^{2}}\right)$, where
$H$ is the Hubble parameter, $a$ is the scale factor and $k$ is
the curvature. It has been found that this model does not only
avoid the causality problem and is phenomenologically viable, but
also naturally solves the coincidence problem \cite{Suwa}. The
energy density of RDE is given by
$\rho_{RDE}=3c^{2}\left(\dot{H}+2H^{2}+\frac{k}{a^{2}}\right)$. In
flat FRW universe, $k=0$ and hence
$\rho_{RDE}=3c^{2}\left(\dot{H}+2H^{2}\right)$.
\\
Interacting DE models have gained immense interest in recent
times. Works in this direction include \cite{setare1},
\cite{setare2}, \cite{setare3}. Interacting RDE was considered in
\cite{Suwa}, where the observational constraints on interacting
RDE were investigated. In this work, we have considered an
interacting RDE. We have reconstructed the Hubble's parameter $H$
under this interaction and subsequently calculated the equation of
state parameter $w$, deceleration parameter $q$ and statefinder
parameters $\{r,s\}$ in terms of redshift $z$. This study deviates
from \cite{Suwa} in the choice of the interaction term. Moreover,
contrary to \cite{Suwa}, we have not considered radiation while
considering interaction and expressed $H$ in terms of redshift
$z$. In \cite{Feng}, statefinder diagnostics of RDE was
investigated without any interaction. Our study deviated from
\cite{Feng} in this regard.
\\\\

\subsection{\bf{Interacting RDE}} The metric of a spatially flat
homogeneous and isotropic universe in FRW model is given by

\begin{equation}
ds^{2}=dt^{2}-a^{2}(t)\left[dr^{2}+r^{2}(d\theta^{2}+sin^{2}\theta
d\phi^{2})\right]
\end{equation}

where $a(t)$ is the scale factor. The Einstein field equations are
given by

\begin{equation}
H^{2}=\frac{1}{3}\rho
\end{equation}
and
\begin{equation}
\dot{H}=-\frac{1}{2}(\rho+p)
\end{equation}

where $\rho$ and $p$ are energy density and isotropic pressure
respectively (choosing $8\pi G=c=1$).\\

The conservation equation is given by

\begin{equation}
\dot{\rho}+3H(\rho+p)=0
\end{equation}

As we are considering interaction between RDE and dark matter, the
conservation equation will take the following form

\begin{equation}
\dot{\rho}_{total}+3H(\rho_{total}+p_{total})=0
\end{equation}

where, $\rho_{total}=\rho_{RDE}+\rho_{m}$ and $p_{total}=p_{RDE}$
(as we are considering pressureless dark matter, $p_{m}=0$). As in
the case of interaction the components do not satisfy the
conservation equation separately, we need to reconstruct the
conservation equation by introducing an interaction term $Q$. It
is important to note that the conservation equations imply that
the interaction term should be a function of a quantity with units
of inverse of time (a first and natural choice can be the Hubble
factor $H$) multiplied with the energy density. Therefore, the
interaction term could be in any of the forms \cite{sheykhi}:
$Q\propto H\rho_{RDE}$, $Q\propto
H\rho_{m}$ and $Q\propto H\rho_{total}$.\\

Considering the interaction term $Q$ as $Q=3H\delta\rho_{m}$,
where $\delta$ is the interaction parameter, the conservation
equation (13) takes the form

\begin{equation}
\dot{\rho}_{RDE}+3H(\rho_{RDE}+p_{RDE})=Q
\end{equation}
and
\begin{equation}
\dot{\rho}_{m}+3H\rho_{m}=-Q
\end{equation}

It should be noted that we are considering pressureless dark
matter. It may be noted that similar choice of the interaction
term has been made in \cite{sheykhi}. If $Q>0$, there is a flow of
energy from dark matter to dark energy \cite{cataldo}.

From equations (3) and (7) we express $p_{RDE}$ under interaction
in a flat FRW universe as

\begin{equation}
p_{RDE}=-\left[(2+3c^{2})\dot{H}+6c^{2}H^{2}+\rho_{mo}a^{-3(1+\delta)}\right]
\end{equation}

Using the energy density and pressure of RDE in (6) we express
$H(z)$ as

\begin{equation}
H(z)^{2}=\frac{c^{2}}{-1+2c^{2}}+B_{0}(1+z)^{-\frac{2}{c}+4c}+\frac{2c(1+z)^{3(1+\delta)}\rho_{mo}}{6+3c(3-4c+3\delta)}
\end{equation}

Using the above form of Hubble's parameter in $p_{RDE}$ and
$\rho_{RDE}$ we get the pressure and energy density of RDE under
interaction. Subsequently we calculate the equation of state
parameter $w_{RDE}=\frac{p_{RDE}}{\rho_{RDE}}$ and plot in figure
1 against $z$ for various values of $c^{2}$.

\begin{figure}
\includegraphics[scale=1.2]{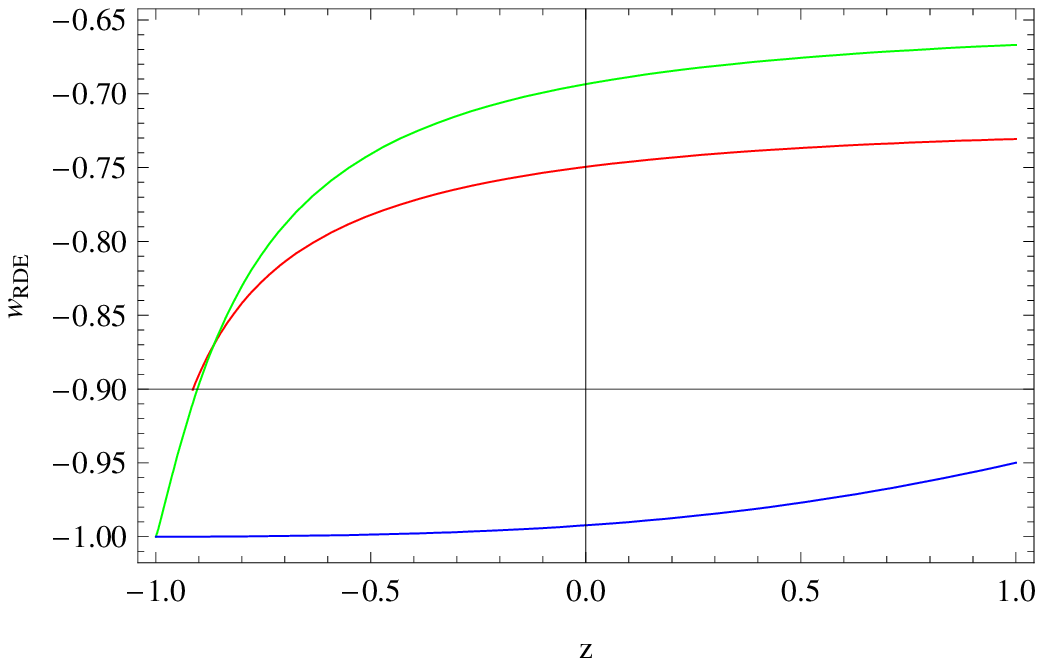}~~~~~\\
\vspace{2mm} \vspace{6mm} Fig. 1 shows the variations of equation
of state parameter $w_{RDE}$ against $z$ for $\delta=0.05$ and
$\rho_{mo}=0.23$. The red, green and blue lines correspond to
$c^{2}=0.7,~0.75$ and $0.8$ respectively. \vspace{6mm}
\end{figure}

\subsection{\bf{Statefinder diagnostics}}
The problem of discriminating different dark energy models is now
emergent. In order to solve this problem, a sensitive and robust
diagnostic for dark energy is a must. The statefinder parameter
pair ${r, s}$ introduced by \cite{Sahni} is proven to be useful
tools for this purpose. The statefinder pair is a `geometrical'
diagnostic in the sense that it is constructed from a space-time
metric directly, and it is more universal than `physical'
variables which depends upon properties of physical fields
describing dark energy, because physical variables are, of course,
model-dependent \cite{Feng}. First, we consider deceleration
parameter $q$

\begin{equation}
q=-\frac{a\ddot{a}}{\dot{a}^{2}}=-1-\frac{\dot{H}}{H^{2}}=-1-\frac{a}{2\tilde{H}^{2}}\frac{d\tilde{H}^{2}}{da}
=-1+\frac{(1+z)}{\tilde{H}^{2}}\frac{d\tilde{H}^{2}}{dz}
\end{equation}

and subsequently we consider the statefinder parameters

\begin{equation}
r = 1+3\frac{\dot{H}}{H^{2}}+\frac{\ddot{H}}{H^{3}}
\end{equation}
and
\begin{equation}
s = -\frac{3H\dot{H}+\ddot{H}}{3H(2\dot{H}+3H^{2})}
\end{equation}

Or, equivalently \cite{ujjal}

\begin{equation}
r=1+\frac{2a}{\tilde{H}^{2}}\frac{d\tilde{H}^{2}}{da}+\frac{a^{2}}{2\tilde{H}^{2}}\frac{d^{2}\tilde{H}^{2}}{da^{2}}
=1-\frac{(1+z)}{\tilde{H}^{2}}\frac{d\tilde{H}^{2}}{dz}+\frac{(1+z)^{2}}{2\tilde{H}^{2}}\frac{d^{2}\tilde{H}^{2}}{dz^{2}}
\end{equation}
and
\begin{equation}
s=-\frac{4a\frac{d\tilde{H}^{2}}{da}+a^{2}\frac{d^{2}\tilde{H}^{2}}{da^{2}}}{3(3\tilde{H}^{2}+a\frac{d\tilde{H}^{2}}{da})}
=-\frac{2(1+z)\frac{d\tilde{H}^{2}}{dz}-(1+z)^{2}\frac{d^{2}\tilde{H}^{2}}{dz^{2}}}{3\left(3\tilde{H}^{2}-(1+z)\frac{d\tilde{H}^{2}}{dz}\right)}
\end{equation}

Using $\tilde{H}=H(z)$ we get under the present interaction the
deceleration parameter as

\begin{equation}
q(z)=-1+\frac{\frac{2(-1+2c^{2})B_{0}(1+z)^{-\frac{2}{c}+4c}}{c}+\frac{6c(1+z)^{3(1+\delta)}(1+\delta)\rho_{mo}}{6+3c(-4c+3(1+\delta))}}{\frac{c^{2}}{2c^{2}-1}+B_{0}(1+z)^{-\frac{2}{c}+4c}+\frac{2c(1+z)^{3(1+\delta)}\rho_{mo}}{6+3c(-4c+3(1+\delta))}}
\end{equation}

The statefinder pair $\{r,s\}$ are obtained as
\begin{equation}
  r=1+\frac{\zeta_{1}}{\zeta_{2}}
\end{equation}

where

\begin{equation}
\begin{array}{c}
\zeta_{1}=3(-1+2c^{2})\times\\
\left\{(-1+2c^{2})(-2+c(-3+4c))B_{0}(1+z)^{-(\frac{2}{c}+3\delta)}(-2+4c^{2}-3c(1+\delta))-3c^{3}(1+z)^{3-4c}\delta(1+\delta)\rho_{mo}\right\}\\
\end{array}
\end{equation}

\begin{equation}
\begin{array}{c}
\zeta_{2}=c^{2}\left\{3(1+z)^{-3\delta}(-1+2c^{2})B_{0}(1+z)^{-\frac{2}{c}}+c^{2}(1+z)^{-4c}\right\}\\
\times(-2+4c^{2}-3c(1+\delta))-2c(-1+2c^{2})(1+z)^{3-4c}\rho_{mo}\\
\end{array}
\end{equation}

\begin{equation}
s=\frac{\xi_{1}}{\xi_{2}}
\end{equation}

where
\begin{equation}
\begin{array}{c}
\xi_{1}=2(-1+2c^{2})\times\\
\left\{(-1+2c^{2})(-2+c(-3+4c))B_{0}(1+z)^{-(\frac{2}{c}+3\delta)}(-2+4c^{2}-3c(1+\delta))-3c^{3}(1+z)^{3-4c}\delta(1+\delta)\rho_{mo}\right\}\\
\end{array}
\end{equation}

and

\begin{equation}
\begin{array}{c}
\xi_{2}=3c(-(-1+2c^{2}))(-2+c(-3+4c))B_{0}(1+z)^{-(\frac{2}{c}+3\delta)}(-2+4c^{2}-3c(1+\delta))+\\
c^{2}(1+z)^{-4c}\left\{3c(1+z)^{-3\delta}(-2+4c^{2}-3c(1+\delta))+2(-1+2c^{2})(1+z)^{3}\delta \rho_{mo}\right\}\\
\end{array}
\end{equation}

The deceleration parameter is plotted against $z$ in figure 2 and
the $\{s-r\}$ trajectory is plotted in figure 3 for different
values of $c^{2}$.

\begin{figure}
\includegraphics[scale=1.0]{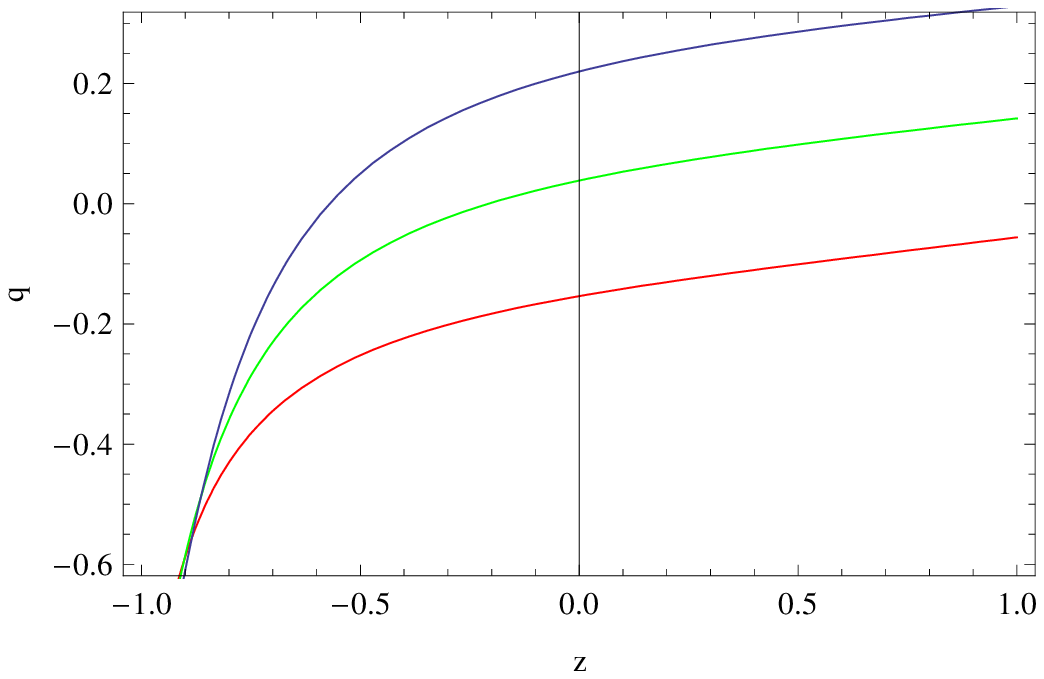}~~~~~\\
\vspace{2mm} \vspace{6mm} Fig. 2 plots the deceleration parameter
$q$ against $z$ for $\delta=0.05$ and $\rho_{mo}=0.23$. The red,
green and blue lines correspond to $c^{2}=0.7,~0.75$ and $0.8$
respectively. \vspace{6mm}
\end{figure}

\begin{figure}
\includegraphics[scale=1.0]{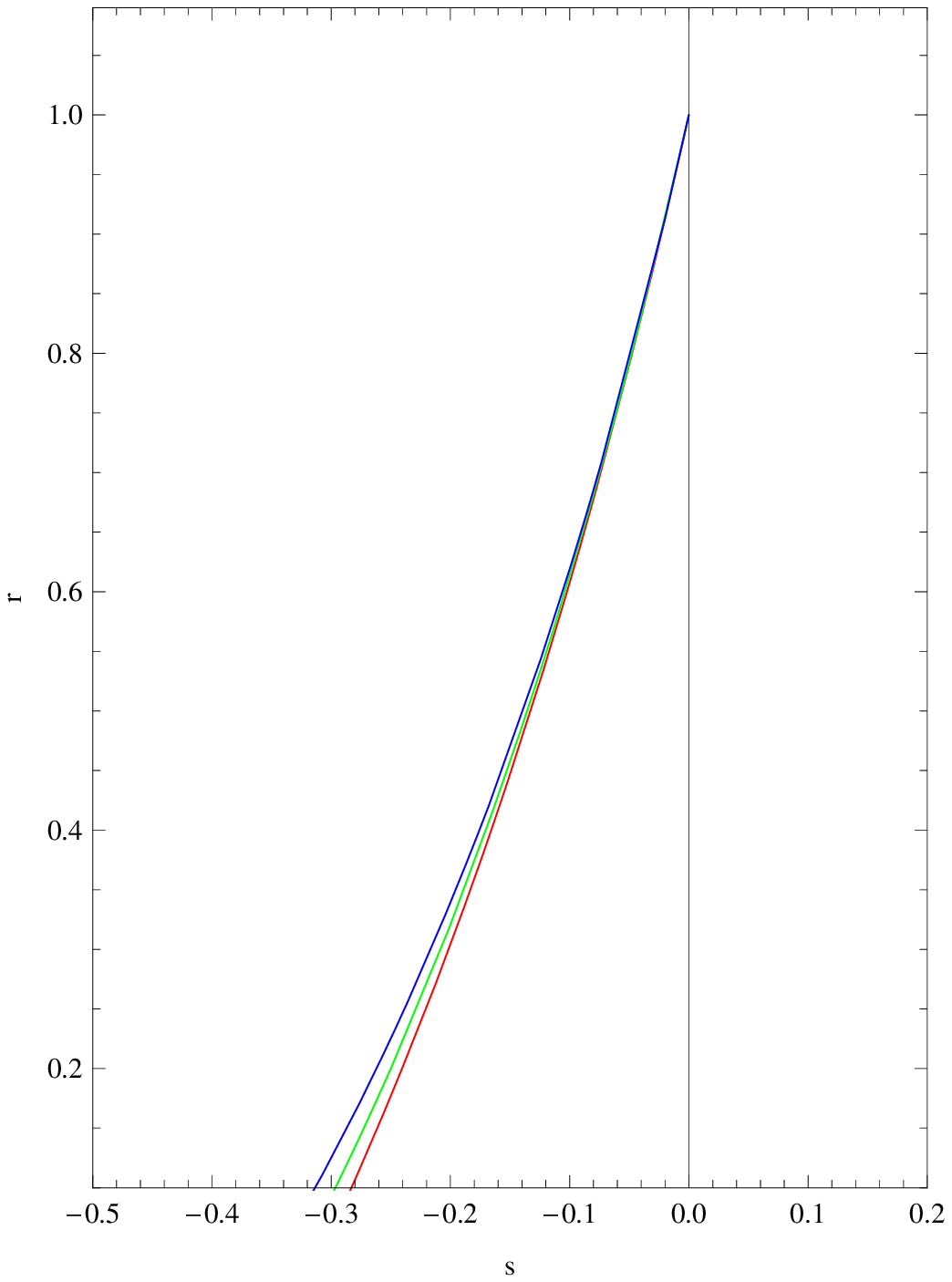}~~~~~\\
\vspace{2mm} \vspace{6mm} Fig. 3 plots the $\{s-r\}$ trajectory
for $\delta=0.05$ and $\rho_{mo}=0.23$. The red, green and blue
lines correspond to $c^{2}=0.7,~0.75$ and $0.8$ respectively.
\vspace{6mm}
\end{figure}

\subsection{\bf{Concluding remarks}}

Considering the interaction between RDE and pressureless dark
matter in a flat Friedman-Robertson-Walker universe we observe the
following:\\
  From figure 1 we observe that the equation of state parameter $w_{RDE}$ is gradually decaying towards $-1$ with
  the evolution of the universe. Throughout the evolution of the universe it is staying above $-1$ and at lower
  redshifts it is tending to $-1$. However, it never crosses the $-1$ boundary. This indicates `quintessence' behavior \cite{saridakis}.
  This observation is consistent with the observation of \cite{chattopadhyay}, where for $c^{2}>1/2$ the equation of state parameter for RDE behaved like
  `quintessence' in presence of dark matter without interaction.\\
  From figure 2 we find that when $c^{2}=0.7$, the deceleration
  parameter $q$ is negative throughout the evolution of the
  universe. However, for $c^{2}=0.75$ and $0.8$, the deceleration
  parameter is transiting from positive to negative side with the
  evolution of the universe. This leads us to conclude that for
  $c^{2}=0.7$ the interacting RDE is giving an ever-accelerating
  universe. However, for the other values of $c^{2}$ the
  interacting RDE is producing the transition from decelerated to
  the current accelerated universe. Moreover, we observe that $q$
  is increasing in the negative side with decrease in the
  redshift. This indicates an increase in the acceleration of the
  universe.\\
  A study of statefinder diagnostics of Ricci dark energy was done by
\cite{Feng}, where it was found that the $\{s-r\}$ trajectory is a
vertical segment, i.e. $s$ is a constant during the evolution of
the universe for a particular choice of $c^{2}$. Figure 3,
suggests a different behaviour of the $\{s-r\}$ trajectory. The
trajectory is mostly confined in the second quadrant of the $s-r$
plane. The spatially flat $\Lambda$CDM (cosmological constant
$\Lambda$ with cold dark matter) scenario corresponds to a fixed
point in the diagram $\{s,r\}|_{\Lambda CDM}=\{0,1\}$. We find
that the $\{s-r\}$ can not cross $\{r=1,s=0\}$. This means that it
can not go beyond $\Lambda$CDM. We also find from figure 3 that
for finite $r$, $s\rightarrow -\infty$. This corresponds to the
dust stage of the universe. Thus, the interacting RDE interpolates
between dust and $\Lambda$CDM stage of the universe.
\\
The study reported in this paper has investigated the behaviors of
the deceleration parameter, equation of state parameters $\{r,s\}$
in presence of interacting Ricci dark energy. As a future study we
propose to investigate current observational constraints on this
model from SNIa, CMB and BAO observations. In particular, it may
be examined whether the present interaction can affect the CMB
constraint. We expect that the future high precision observation
data may enlighten Ricci dark energy further and may reveal some
significant features of the underlying theory of dark energy.
\\
\subsection{\bf{Acknowledgment}}
The author is thankful to the anonymous reviewer for thoughtful
and constructive suggestions to enhance the quality of the paper.
The author sincerely acknowledges the Inter-University Center for
Astronomy and Astrophysics, Pune, which provided Visiting
Associateship to the author.
\\

\end{document}